\documentstyle[12pt,thmsa,sw20aip]{article}

\input{tcilatex}
\begin{document}

\title{A comment on a proposal to use neutrons to reveal gravitomagnetic effects on
Earth}
\author{A. Tartaglia, M.L. Ruggiero \\
Dip. Fisica, Politecnico, and INFN, Torino, Italy\\
e-mail: tartaglia@polito.it ; ruggierom@polito.it}
\maketitle

\begin{abstract}
Arguments are presented which show that the recent proposals of using
neutrons to test gravitomagnetic effects on Earth are impracticable.
\end{abstract}

In a recent paper by Iorio, appeared on these archives \cite{iorio}, an
interesting proposal is put forth: to use neutron interferometry to reveal
gravitomagnetic effects originated by a rapidly rotating laboratory scale
spherical mass. The starting point of this proposal is in the formula that
gives the difference in revolution times for a pair of counter-orbiting
objects: 
\begin{equation}
\Delta T=4\pi \frac{a}{c}  \label{uno}
\end{equation}
where $a=J/Mc$ is the ratio between the angular momentum and the product of
the mass times the speed of light for the central spinning body.

Formula (\ref{uno}) is indeed appealing since the numerical value it gives
for the whole Earth is a good $10^{-7}$ s, furthermore it is independent
from the radius of the circular trajectories and from Newton's constant $G$.
For a reasonable sphere in a laboratory it could be $\Delta T\sim $ $%
10^{-14} $ s \cite{tart1}. Unfortunately however (\ref{uno}) holds only for 
{\it freely} orbiting objects \cite{mashhoon}. In other words the two
circular trajectories must be geodesic.

When the trajectory is not geodesic the difference in revolution times for
objects moving with the same local speed would be \cite{tart2} 
\begin{equation}
\Delta T=12\pi \frac{GM}{c^{2}r}\frac{a}{c}  \label{due}
\end{equation}
which depends on the radius of the circle $r$ and on $G.$ Numbers are in
general much smaller than before.

Using the data suggested by Iorio in his paper one would obtain (taking a $1$
- meter radius for the trajectory )

\[
\Delta T\sim 10^{-43}\text{ }s 
\]

$\allowbreak $

and a phase shift

\[
\Delta \Phi _{\pi }\sim \allowbreak \allowbreak 10^{-28}\text{ }rad 
\]

too low to be revealed.

If one wanted actually to apply formula (\ref{uno}) with neutrons, one had
to have them freely orbiting around the spinning mass. Suppose the central
mass is in the order of $100$ kg, a ''satellite'' on a circular orbit $2$ m
wide in diameter would move at a speed in the order of $10^{-4}$ m/s. This
speed would correspond to extremely cold neutrons whose wave length would be 
$\lambda \sim $ $10^{-3}$ m. We are not aware of a present capability to
produce beams of such extra long wave length neutrons.

Using quantum particles to measure gravitomagnetic effects is an exciting
and intriguing idea but apparently we have not yet found the right way.

\end{document}